\documentclass[aps,prl,twocolumn,groupedaddress]{revtex4}
\usepackage{graphicx}
\usepackage{dcolumn}
\usepackage{bm}
\usepackage{amssymb}

\def\lsim{\mathrel{\mathstrut\smash{\ooalign{\raise2.5pt\hbox{$<$}\cr\lower2.5pt\hbox{$\sim$}}}}}
\def\gsim{\mathrel{\mathstrut\smash{\ooalign{\raise2.5pt\hbox{$>$}\cr\lower2.5pt\hbox{$\sim$}}}}}

\def\be{\begin{equation}}
\def\ee{\end{equation}}

\begin{document}

\title{Weighing Neutrinos with Galaxy Cluster Surveys}

\author{Sheng Wang$^{1,2}$, Zolt\'{a}n Haiman$^{3}$, Wayne Hu$^{4}$,
Justin Khoury$^{5}$, and Morgan May$^{1}$}

\affiliation{$^1$Brookhaven National Laboratory, Upton, NY 11973--5000, USA \\
$^2$Department of Physics, Columbia University, New York, NY 10027, USA \\
$^3$Department of Astronomy, Columbia University, New York, NY 10027, USA \\
$^4$Kavli Institute for Cosmological Physics, Department of Astronomy and Astrophysics,
Enrico Fermi Institute, University of Chicago, Chicago, IL 60637, USA \\
$^5$Center for Theoretical Physics, Massachusetts Institute of Technology, Cambridge, MA 02139, USA}

\leftline{MIT-CTP 3644}

\begin{abstract}

Large future galaxy cluster surveys, combined with cosmic microwave
background observations, can achieve a high sensitivity to the masses
of cosmologically important neutrinos. We show that a weak lensing
selected sample of $\gsim 100,000$ clusters could tighten the current upper
bound on the sum of masses of neutrino species by an order of
magnitude, to a level of 0.03 eV. Since this statistical sensitivity is
below the best existing lower limit on the mass of at least one
neutrino species, a future detection is likely, provided that
systematic errors can be controlled to a similar level.

\end{abstract}

%\pacs{}

\maketitle

\begin{figure}
\resizebox{80mm}{!}{\includegraphics{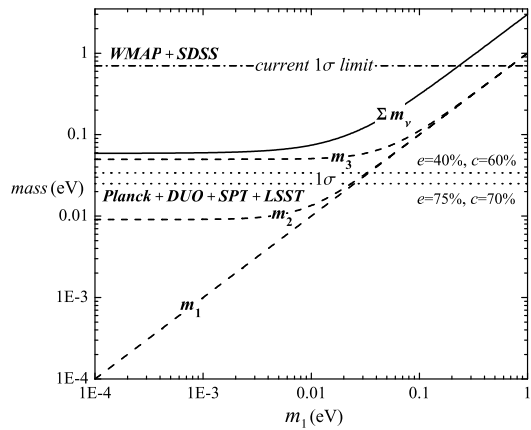}}
\caption{\label{fig:numass} The total mass and three mass eigenvalues
of neutrinos as a function of the unknown smallest mass
$m_1$~\cite{bb02}. A normal hierarchy is assumed.  In our analysis, we
consider the scenario of three families of neutrinos with one massive
and two massless. The upper dotted line shows the current most
robust $1\sigma$ limit~\cite{tegmark04}.
The lower dotted lines show our predicted $1\sigma$
constraints, for two different sets of weak lensing selection
efficiency and completeness, from combining all three cluster surveys
and Planck.}
\end{figure}

Recent experiments have placed both stringent lower and upper bounds
on the masses of neutrinos.  The lower bounds derive from neutrino
oscillations experiments, which measure the difference in the squared
masses of the coupled species.  Atmospheric neutrino oscillations
showed that at least one neutrino species has a mass $\gsim
0.05$~eV~\cite{fukuda98}, implying that neutrinos make a
non--negligible contribution to dark matter. Solar neutrino
oscillations showed smaller mass splittings ($\sim 0.008$~eV
\cite{abaz03}) (see Fig.~\ref{fig:numass} for the allowed masses of
individual species).

The upper bounds come from cosmological large scale structure
measurements, which are sensitive to the sum of the masses of all
neutrinos.  By combining CMB anisotropies from the Wilkinson
Microwave Anisotropy Probe (WMAP)~\cite{spergel03} and the clustering
properties of galaxies in the Sloan Digital Sky Survey (SDSS), with no
prior assumptions about the amplitude of galaxy bias, $\sum m_\nu <
1.7$~eV at 95\%~CL was obtained~\cite{tegmark04}. The limit can be
tightened to 0.42 eV~\cite{seljak04} by incorporating weak lensing,
Ly$\alpha$ forest and Supernovae data, but at the expense of
introducing possible new systematics~\footnote{Although we note that
prior work on neutrinos have assumed a fixed equation of state $w=-1$
for the dark energy, whereas we allow $w$ to vary.  Our results on the
neutrino mass would improve by a factor of 1.5 for DUO and SPT, a
factor of 2 for LSST if we had also fixed $w=-1$.}.

The above bounds have narrowed the allowed range for the sum of the
masses $\sum m_\nu$ to within an order of magnitude.  In this {\it
Letter}, we discuss the prospect of closing the gap, utilizing large,
future surveys of galaxy clusters.  Such surveys, covering large
fractions of the sky to impressive depths, are being planned in
several wavelength bands.  There are several
advantages of such a cluster sample. First, clusters correspond to
massive dark matter potential wells, whose abundance and spatial
distribution are dictated by gravity alone, and have been simulated to
high accuracy~\cite{jenkins01}.

Second, galaxy clusters are highly clustered relative to the mass
distribution. This bias can, at least principle, be determined
from simulations, and it increases the signal--to--noise in power
measurements by a factor of $\sim 10-100$ (for the relevant cluster
masses below).  Third, a galaxy cluster survey delivers several
observables. As we will find below, the power spectrum $P_c(k)$ and
abundance evolution $dN/dz$ provide complementary constraints: the
free--streaming of neutrinos suppresses fluctuation power on small
scales, affecting the shape of $P_c(k)$, and also the growth rate of
perturbations in the linear regime $g(z)$, affecting $dN/dz$.  The
main limitation of using galaxy clusters is that the relation between
cluster mass and the actual observables needs to be known to high
precision, at least statistically.  As we discuss below, this should
be achievable using empirical calibrations and numerical simulations.

{\em Neutrino Signature.} --- Massive neutrinos cluster on very large
scales, but free--stream out of small--scale dark matter potential
wells and thus suppress density fluctuations on small scales. In
addition, the evolution of perturbations is no longer independent of
scale. The late--time evolution of perturbations in a cosmology with
cold dark matter (CDM), baryons and massive neutrinos, accounting for
all post--recombination effects, can be accurately treated using a
product of a scale--dependent growth function (Eq.~25 in~\cite{huei98})
and a time--independent transfer function, reflecting conditions at the
drag epoch. The latter is obtained numerically from CMBFAST~\cite{seljak96}
and includes acoustic baryonic features, which contain information
comparable to the broader features at the matter--radiation equality
and the sound horizon~\cite{huha03}.

{\em Cluster Abundance and Distribution.} --- Since our analysis
closely follows the treatment of~\cite{us}, here we only outline the
methods and refer the reader to~\cite{us} for details.  We use the
results of numerical simulations from Jenkins {\it et
al.}~\cite{jenkins01} for the differential comoving number density
$dN/dM(M,z)$ of clusters of total mass $M$ at redshift $z$ (the
fitting formula in their Eq.~B3). Correlations in the spatial
distribution of clusters are described by the power spectrum,
$P_c(k)$, taking into account redshift space
distortions~\cite{kaiser87} and bias~\cite{sheth99}.

{\em Future Cluster Surveys.} --- Galaxy clusters can be identified
and cataloged using several different observational signatures. We
consider three methods that use the X--ray emission of the hot cluster
gas, the up--scattering of CMB photons in energy by this gas
(Sunyaev--Zel'dovich effect, hereafter SZE), or the distortion of the
images of background galaxies by weak gravitational lensing (WL).  To
determine the detection mass limit $M_{\rm min}(z)$ in each case, we
adopt a flux limited X--ray survey (guided by plans of the Dark
Universe Observatory, DUO~\cite{duo}, yielding a total of $\approx
11,500$ clusters in 6,150 deg$^2$), a temperature--decrement limited
SZE survey (as planned with the South Pole Telescope, SPT~\cite{spt},
yielding $\approx 20,000$ clusters in 4,000 deg$^2$), and a
shear--limited WL survey (planned with the Large Synoptic Survey
Telescope, LSST~\cite{lsst}, yielding $\approx 200,000$ clusters in
18,000 deg$^2$).  We have additionally imposed a floor for the minimum
mass of $10^{14} h^{-1} M_{\odot}$ in the X--ray and SZE surveys,
since less massive halos correspond to small clusters or groups in
which these two observables are subject to strong non--gravitational
effects. If we imposed the same minimum mass floor for LSST, it
would yield $\approx 143,000$ clusters. The mean redshifts are 0.36
for DUO, 0.58 for SPT and 0.43 for LSST (see~\cite{us}, Fig. 1 for the
redshift distribution of clusters).  To parameterize these effects
for the clusters above the detection threshold, we introduce three
additional non--cosmological parameters in predicting both the X--ray
flux and the SZ decrement, of the form $f_x(z) \propto A_x M^{\beta_x}
(1+z)^{\gamma_x}$ and $f_{sz} \propto A_{sz}
M^{\beta_{sz}}(1+z)^{\gamma_{sz}}$, respectively.  Deviations
from these power--law scalings are not captured by our
parameterization and would have to be addressed by allowing more
general forms of cluster structure and evolution (see discussion
below).  For the WL survey, we assume that clusters (even at
low--mass) produce shear signals with a well--defined mass--shear
relation and do not introduce analogous free parameters. The main
systematic error for WL surveys comes from false detections and
incompleteness (discussed below).

{\em Fisher Matrix Forecasts.} --- We estimate the constraints on
cosmological parameters using the Fisher matrix formalism that has
become a standard tool~\cite{tegmark97}, allowing easy combination of
different independent data sets and/or methods, by summing individual
matrices.  The Fisher matrix for the redshift distribution and power
spectrum are described in detail in~\cite{us}. The Fisher matrix for
the temperature and E--mode polarization anisotropy of the cosmic
microwave background (CMB) is constructed as given
by~\cite{zaldarriaga96}. In this paper, the covariance matrix of WMAP
is from the first year data, and the expected parameters of the Planck
satellite~\cite{planck} are listed in Table~I of~\cite{us}.

The Fisher formalism requires a fiducial model around which variations
are considered. For the cosmological parameters, we adopt a
spatially--flat, CDM model, dominated by a cosmological constant. 
The set of parameters included in our analysis is \{$\Omega_b h^2$,
$\Omega_m h^2$, $\Omega_{\nu} h^2$, $\Omega_{\rm DE}$, $w$, $n_s$,
$\sigma_8$, $\tau$\}, where all the symbols have their standard
meaning. The values are adopted from recent measurements by WMAP, as
summarized in Table~1 of~\cite{spergel03}: \{0.024, 0.14, 0, 0.73, -1,
1, 0.9, 0.17\}.  We find that choosing $\Omega_{\nu} h^2=0.00058$ --
the current lower limit -- would not change our results below.

Our analysis takes into account the dominant systematic errors for the
three surveys. For X--ray and SZE surveys, these come from
uncertainties in the mass--observable relation due to structure and
evolution of clusters. As proposed in~\cite{majumdar03,majumdar04},
however, one can use cluster surveys not only as a source of
cosmological information, but also to constrain the mass--observable
relation, thereby making the survey self--calibrating.  To account for
this, we include the parameters of the mass--observable relation in
the Fisher matrix analysis: $A_x$ ($=10^{-4.159}$), $\beta_x$
($=1.807$) and $\gamma_x$ ($=0$) for X--ray survey; $A_{sz}$
($=10^{8.9}$), $\beta_{sz}$ ($=1.68$) and $\gamma_{sz}$ ($=0$) for SZE
survey, where numbers in parentheses indicate the fiducial values for
these parameters~\cite{majumdar04}.

The main systematic limitation for WL surveys comes from false
detections and incompleteness. In our analysis of the LSST--like WL
survey, we used a constant shear S/N threshold to select
clusters. False detections or missing clusters result from statistical
fluctuations in these ellipticities and from projections of physical
structures along the line of sight. Several
papers~\cite{hennawi04,hamana03} have done a comprehensive study of
mass--selected clusters using N--body simulations. They point out that
because the simulations depend only on gravity, the expected cluster
distribution, including false detections and missing clusters, can be
reliably calculated for any cosmological model.  Thus, false
detections and missing clusters can be accounted for, and their
presence serves only to increase the statistical error.  We here adopt
the efficiency $e=40\%$ and the completeness $c=60\%$ for a 4.5
standard deviation detection threshold~\cite{hamana03}. To account for
this, we multiply our parameter error bars for the WL survey by a
correction factor of $\sqrt{[(1/e-1)+1/e]/c} \sim 2.6$.

{\em Results and Discussion.} --- Table~\ref{tab:clustersandcmb} shows
the neutrino mass constraints from our Fisher matrix analysis for CMB,
and three types of cluster surveys, including self--calibration for
X--ray and SZE surveys, as well as false detection and completeness
for the WL survey.  The constraint from the forecast for Planck alone
is $\sigma (\sum m_{\nu})=0.23$~eV~\cite{eisenstein98}.

For all three cluster surveys, the power spectrum, $P_c(k)$, is a much
more sensitive probe of the neutrino mass than the counts, $dN/dz$, as
expected, and, by itself, yields a constraint of $\sigma (\sum
m_{\nu})\sim 1$~eV. Combining with WMAP and Planck improves this by
factors of $\approx 5$ and $\approx 10$, respectively.  Similarly,
combining $P_c(k)$ with $dN/dz$ yields a factor of 2 improvement for
DUO and SPT, but only a modest change for LSST. (This is because LSST
measures a smaller fraction of clusters at higher redshift, as seen
from Fig.~1 of~\cite{us}, therefore making $dN/dz$ less relevant for
this survey.) Hence we find that each cluster survey, combined with
WMAP, yields $\sigma (\sum m_{\nu})\sim 0.1$~eV.  When combined with
Planck, each survey gives $\sigma (\sum m_{\nu})=0.04-0.07$~eV, very
near the interesting limit of $0.05$~eV from the Super--Kamiokande
atmospheric neutrino experiment~\cite{fukuda98}.

\begin{table}
\caption{\label{tab:clustersandcmb}Estimated Constraints on Total Mass
of Neutrinos (in unit of eV). The DUO--like and SPT--like surveys are
self--calibrated by including three non--cosmological parameters in
our analysis.  The errors in the LSST--like survey incorporate
calibration with numerical simulations, assuming an efficiency of
$e=40\%$ and completeness $c=60\%$; numbers in parentheses assume
$e=75\%$ and $c=70\%$.  'W' and 'P' stand for WMAP and Planck
respectively.}
\begin{ruledtabular}
\begin{tabular}{|c|ccc|}
&DUO&SPT&LSST\\
&\footnotesize{(6,150 deg$^2$)}&\footnotesize{(4,000 deg$^2$)}&\footnotesize{(18,000 deg$^2$)}\\
\hline
$P_c(k)$&1.4&1.1&0.71~(0.42)\\
%$dN/dz$&3.1&11.&11.~(6.8)\\
$P_c(k)$+$dN/dz$&0.70&0.72&0.53~(0.32)\\
$P_c(k)$+$C_{\ell}$(W)&0.22&0.20&0.15~(0.11)\\
$P_c(k)$+$C_{\ell}$(W)+$dN/dz$&0.16&0.15&0.12~(0.10)\\
$P_c(k)$+$C_{\ell}$(P)&0.11&0.10&0.086~(0.061)\\
$P_c(k)$+$C_{\ell}$(P)+$dN/dz$&0.071&0.062&0.040~(0.027)\\
\hline
\multicolumn{4}{|l|}{DUO+SPT+LSST+Planck~~~~~~~~~~0.034~(0.025)}\\
\end{tabular}
\end{ruledtabular}
\end{table}

For the X--ray and SZE surveys, one can ask how much could be gained
by completely eliminating the systematics due to cluster structure and
evolution. We can estimate this by dropping the corresponding
parameters, $A_x$, $\beta_x$, etc. from the Fisher matrix. When
combined with Planck, each survey would give $\sigma (\sum
m_{\nu})=0.02-0.03$~eV, which is well below the Super--K bound.  While
a complete elimination of systematic errors is unrealistic, it is
pointed out in~\cite{hu03} that including the additional information
from the shape of the mass function ($dN/dM$) allows one to largely
eliminate ambiguities caused by a wide range of possible evolutions of
the mass--observable relation.  Here we consider reducing the
degradation of constraints due to self--calibration by combining DUO
and SPT together (i.e. statistically; no overlap of the survey areas
is then needed). Once again with Planck, we now find $\sigma (\sum
m_{\nu})\approx 0.054$~eV, back below the lower limit of $\sum
m_{\nu}\approx 0.058$~eV implied by the combination of Super--K and
solar oscillation experiments.

For the WL survey, it has been demonstrated~\cite{hennawi04} that a
tomographic analysis can significantly reduce
projections. Also,~\cite{gladders01} has shown that red galaxies can
be used to selected clusters optically with a $<5\%$ false detection
rate.  LSST will contain both lensing and optical observations of
clusters, and projection effects and false detections can be removed
rather than statistically subtracted.  A detailed study of these and
other possible enhancements, such as applying a series (rather than a
single) smoothing filter to the WL shear maps, which were not
considered by~\cite{hamana03}, is needed. If a completeness of 70\%
and efficiency of 75\% can be attained without introducing a new
systematic bias, then a 0.027eV sensitivity of neutrinos from LSST
will be achievable.  In Table~\ref{tab:clustersandcmb}, numbers in
parentheses in the LSST column correspond to $c=70$\% and $e=75$\%.

We have checked the effects of several other systematics, and found
them to be small:
possible redshift evolution in the bias $b(z)$~\cite{us,huha03};
scatter in the mass-observable relation~\cite{limahu05}; baryon
cooling altering the density profile of clusters~\cite{white04}.
One important systematic error is the uncertainty of photometric
redshifts. The redshift uncertainty of one single red galaxy is
0.03 for low redshifts and approximately twice worse for high
redshifts~\cite{padma04}. On the other hand, a cluster of mass
as large as $10^{14} h^{-1} M_{\odot}$ should have at least 25
red galaxies~\cite{cosh02}, and the redshift uncertainty of 
each cluster as a whole would be reduced after averaging  
the redshifts of the constituent galaxies.
We find that
an error of $\Delta z=0.03$ would degrade the constraint on $\sum
m_{\nu}$ from $P_c(k)$ alone by a factor of $\sim 2$, as modes with a
relatively large $k_{\parallel}$ will be swamped by shot noise and
therefore give no leverage~\cite{seo03}.  However, after combining
with $dN/dz$ and CMB, the effect is much smaller: a $\approx 20\%$
increase on the final errors.

There are other cosmological parameters that could be added in
our analysis, for example, running of the scalar spectral index
($dn_s/dk$) or an evolution of the dark energy equation of state
($dw/dz$).  However, the constraint on neutrino mass derives from the
evolving scale--dependence of the growth function; an unevolving
curvature in $P_c(k)$ as a function of scale, or a smooth evolution of
the equation of state, can not mimic such a scale--dependent
evolution.  As an example, we have explicitly verified that adding an
additional free parameter, $\alpha\equiv d n_s/d\ln
k$~\cite{spergel03}, increases our final error on the neutrino mass by
less than 2 percent.

To conclude, taking into account self--calibration for X--ray and SZE
surveys, as well as completeness and efficiency for WL surveys, we
find a combined constraint for DUO + SPT + LSST + Planck of $\sigma
(\sum m_{\nu})= 0.034$~eV. This implies at least a 1.7$\sigma$
detection of neutrino dark matter; improving WL selection efficiency
and completeness would increase the significance to 2.3$\sigma$ (see
Table 1).

We consider how our constraints could be improved by a measurement of
the power spectrum from the Ly$\alpha$ forest. Using the
one--dimensional power spectrum $k_\parallel P_{1D}(k_\parallel)$ as
an observable, measured at the single value of line-of-sight
wavenumber $k_{\parallel}=1 h$~Mpc$^{-1}$, we find that our
constraints on $\sum m_{\nu}$ for all cluster surveys improve by a
factor of 2 if the 1D power spectrum can be measured to an accuracy,
including control of systematic errors, better than $\Delta P/P \sim
1$\% (note this accuracy would allow a more modest improvement of 30\%
if a $w=-1$ prior was adopted). This yields an improved prediction of
a $\gsim 2\sigma$ detection of the neutrino mass.  We find that adding
constraints from 2,000 SNe between $0<z<1.7$, whose magnitudes are
measured to an accuracy of 0.15 mag, following the Fisher matrix
analysis of~\cite{linder03}, yields only a modest 20\% improvement on
$\sum m_{\nu}$.
An additional complementary probe is tomographic cosmic shear
statistics. Comparable bounds on $\sum m_{\nu}$ were found for an
LSST-like survey combined with Planck~\cite{song04}.

In conclusion, our results suggest that detection of the effect of
neutrino dark matter is likely to be possible.  Systematic errors
such as projections and false detections in the WL survey may be
controlled using self--calibration and comparisons to numerical
simulations. However, we emphasize that additional biases will be
introduced by selection effects in real surveys.  These must be
carefully studied and controlled in analyzing any future survey data.
Finally, we note that at the current lower limit on their mass,
neglecting neutrinos from the type of analysis we outlined would
result in a bias of other cosmological parameters. We find, for
example, that $w\approx -0.95$ would be inferred if neutrinos are
ignored in an $w=-1$ universe.  Hence, it is important to include
neutrinos in any analysis that aims to derive dark energy parameters
to $\sim 1\%$ precision.

We thank G.~Holder, J.~Hennawi, L.~Knox, D.~Spergel and J.A.~Tyson for insightful
discussions.  This work is supported in part by the U.S. Department of
Energy under Contract No. DE-AC02-98CH10886, and by DE-FC02-94ER40818
(J.K.), and the Packard Foundation (W.H.).

\end{document}